\documentclass[aps,pr,preprint,floatfix]{revtex4}
\usepackage{graphicx,amsmath,units,xspace,amssymb}

\begin{document}
\title{Visualizing kinetic pathways of homogeneous nucleation in colloidal crystallization}


\author{Peng Tan$^1$, Ning Xu$^2$ and Lei Xu$^{1*}$}
\affiliation{1. Department of Physics, The Chinese University of
Hong Kong, Hong Kong, China\\2. Department of Physics, Hefei
National Laboratory for Physical Sciences at the Microscale, and CAS
Key Laboratory of Soft Matter Chemistry, University of Science and
Technology of China, Hefei, China}
\date{\today}

\begin{abstract}
When a system undergoes a transition from a liquid to a solid phase,
it passes through multiple intermediate structures before reaching
the final state. However, our knowledge on the exact pathways of
this process is limited, mainly due to the difficulty of realizing
direct observations. Here, we experimentally study the evolution of
symmetry and density for various colloidal systems during
liquid-to-solid phase transitions, and visualize kinetic pathways
with single-particle resolution. We observe the formation of
relatively-ordered precursor structures with different symmetries,
which then convert into metastable solids. During this conversion,
two major cross-symmetry pathways always occur, regardless of the
final state and the interaction potential. In addition, we find a
broad decoupling of density variation and symmetry development, and
discover that nucleation rarely starts from the densest regions.
These findings hold for all our samples, suggesting the possibility
of finding a unified picture for the complex crystallization
kinetics in colloidal systems.
\end{abstract}

\maketitle

\noindent\textbf{Introduction}\\
Crystallization is an important process in condensed matter physics
and materials science. As liquid changes into solid, evolutions in
both symmetry and density take place. However, the exact kinetic
pathways across the liquid-to-solid transition remain as a focus of
debate, with no unified picture up to date. Although the classical
nucleation theory (CNT) provides a nice framework for understanding
crystallization, it has been well recognized that it does not
properly describe all aspects of the nucleation process. In the CNT
description, the crystallization kinetics proceeds as nucleation and
growth of nuclei: small crystals, which have the same symmetry and
density as the stable solid, directly emerge from liquid through
spontaneous density fluctuation. By contrast, however, previous
studies find rather different results: multiple metastable solids
that may differ from the stable solid are observed, with one
dominant component prevailing in the nucleation process
\cite{ostpc1897, alexprl78}. More specifically, the body-centered
cubic (BCC) symmetry dominates the early crystallization in
Lennard-Jones and other soft-potential systems
\cite{woldeprl95,woldejcp96,shenprl96,frenkeljpcm02,Moroniprl05,tanakasm};
while the random hexagonal close-packed (RHCP) structure makes the
dominant metastable solid in hard-sphere systems
\cite{puseynature86, puseyprl89, zhunature97, gasserscience01,
Auernature01} (RHCP is a mixture of face-centered cubic (FCC) and
hexagonal close-packed (HCP) structures). The kinetic origin of
these metastable solids, especially of the dominant component, is
currently an outstanding issue requiring experimental elucidation.

Correspondingly, the density variation accompanying such symmetry
transformation remains unclear. The traditional view of CNT
indicates a simultaneous density and symmetry variation within one
step; while recent simulations suggest the possibility of either a
two-step variation where densification precedes order construction
\cite{Woldescience}, or a gradual transition in both symmetry and
density driven by the bond order fluctuation
\cite{tanakapnas,tanakasr}. The exact density-symmetry correlation
depends on the kinetic transformations among different intermediate
structures which can be best clarified with single-particle level
experiments.

The missing of kinetic information also prevents a clear
understanding on the liquid state immediately before
crystallization. Unlike the conventional picture of a uniformly
disordered liquid state, numerical simulations suggest the existence
of relatively-dense or relatively-ordered liquid structures serving
as precursors of nuclei \cite{Schillingprl, Tothprl, lechnerprl,
tanakapnas, tanakasr, Woldescience, Lutskoprl}, which may account
for an intermediate crystallization step
\cite{Martinpre03,SchopePRL,Schopejcp,Savageprl}. However, the
entire picture is far from clear. How does liquid pass through the
precursor state to become metastable solids remains a big mystery --
even the direct observation of precursors in three-dimensional (3D)
real space is still missing. To tackle this mystery, illustrating
the crystallization kinetics again plays an essential role.

In this study, we report single-particle level measurements on the
kinetics of 3D colloidal crystallization, for both symmetry and
density transformations. To dig out the universal kinetics,
extensive studies across different final states and interaction
potentials are performed. Colloidal systems are used due to their
similar phase-transition behaviors as atomic and molecular systems,
while their large particle sizes enabling single-particle level
visualization
\cite{larsennature97,weeksscience2000,gasserscience01,andersonnature2002,schallscience2004,yodhscience2005,salvagescience06,lunature2008,yilongscience12}.
We directly observe relatively-ordered liquid structures behaving as
precursors, out of which small nuclei emerge. These precursors
exhibit local orders close to HCP, BCC and FCC symmetries, and
subsequently convert into these metastable solids. During this
precursor-to-solid conversion, interestingly, two major
cross-symmetry pathways are universally observed: (1) HCP-like
precursors to BCC metastable solid and (2) HCP-like precursors to
FCC metastable solid. These major pathways could explain the kinetic
origin of the dominant metastable solid. In the density evolution,
we surprisingly discover a strong decoupling between density
variation and symmetry development, and demonstrate that initial
nucleation rarely starts from the densest regions of our samples.

Our system is made up of poly(methyl methacrylate) colloids with
diameter $\sigma=2.2 \mu m$ and polydispersity less than 2.5$\%$,
suspended in a density and refractive index matched solvent. The
particles carry charges and form Wigner crystals at low
concentrations \cite{LeunissenNature2005, Pengprl2012}. We fix the
sample concentration at 17\% which avoids particle-particle contact
and ensures soft-repulsive potential (see ``Methods''). By tuning
the Debye screening length, $\kappa^{-1}$, and the surface charge,
$Z$, we can obtain BCC-stable and FCC-stable crystals
respectively\cite{LeunissenNature2005, Hynninen}. The phase diagram
is shown in the Supplementary Information (SI, Fig.SI-1A). The
adjustable stable states offer the opportunity to study kinetic
pathways for various transition situations. We shear melt colloidal
crystals and record the re-crystallization process immediately after
agitation, at the supercooling of $\Delta T = T_m - T = 0.2T_m\sim
0.4 T_{m}$, with $T_m$ being the melting temperature (see
``Methods'').\\

\noindent\textbf{Three symmetries in precursors and nuclei}\\
By inspecting the local bond order parameters (shown later), we can
specify highly-ordered nuclei particles and relatively-ordered
liquid particles (in comparison to normal liquid particles), as
shown respectively by the dark-brown and light-brown spheres in
Fig.1A-C (movie S1): at the beginning of nucleation in Fig.1A, small
nuclei (dark brown) emerge from relatively-ordered liquid structures
(light brown) which serve as \emph{precursors} of solid. Moreover,
such precursors are continuously created in the subsequent
development, as demonstrated by Fig.1B around the stage of critical
nuclei size and Fig.1C for the post-critical stage. Precursors wrap
the nuclei like a layer of coating and account for an intermediate
step in the liquid-solid transition \cite{Schillingprl, lechnerprl,
tanakapnas, tanakasr,Woldescience,Lutskoprl}.

To characterize the local order of any particle $i$, we measure its
local bond order parameters, $q_{l}(i)=(\frac{4
\pi}{2l+1}\Sigma_{m=-l}^{m=l}|q_{l,m}(i)|^2)^{1/2}$,
$W_l(i)=\Sigma_{m_{1},m_{2},m_{3}=0}^{l}(\begin{array}{ccc}l & l & l
\\m_{1} & m_{2} & m_{3}
\end{array})\frac{q_{l,m_{1}}(i)q_{l,m_{2}}(i)q_{l,m_{3}}(i)}{|q_{l}(i)|^{3}}$,
and the coarse grained bond order parameters,
$\overline{q}_l(i)=(\frac{4\pi}{2l+1}\Sigma_{m=-l}^{m=l}|\overline{q}_{l,m}(i)|^2)^{1/2}$
(see SI for more information) \cite{steinhardtprb83, woldejcp99,
lechnerjcp08}. Good statistics is obtained by locating over
$2\times10^{6}$ particles at 50 different regions within the same
sample, in the early crystallization stage (solid fraction $\sim
5\%$). A clear distinction between nuclei and liquid is demonstrated
by the number distribution of all particles in the
$\overline{q}_6$-$\overline{q}_4$ plane in Fig.1D: the upper group
with larger $\overline{q}_6$ is composed by nuclei particles; while
the lower group with smaller $\overline{q}_6$ is by liquid
particles. Within the liquid group, we further pick out the
particles with relatively high bond order, $\overline{q}_6 > 0.27$,
as the light-brown precursors shown in Fig.1A-C. According to the
two-step nucleation theory, precursors are dense structures with a
symmetry lower than solid \cite{Tothprl, Schillingprl, SchopePRL,
Lutskoprl}. Our precursors are therefore defined as the particles
with relatively high local order ($\overline{q}_6 > 0.27$) but below
the level of solid structures (solid bond number $\xi<7$
\cite{woldejcp99}), in agreement with Kawasaki \emph{et al}
\cite{tanakapnas}. With this definition ($\overline{q}_6 > 0.27$ and
$\xi<7$), the precursors also have densities higher than normal
liquid (see Fig.SI-4D in SI), which is consistent with previous
simulations \cite{Tothprl, Schillingprl, SchopePRL, Lutskoprl}. Note
that we can further vary the threshold value of 0.27 to numbers
between 0.27 to 0.30, and obtain qualitatively similar results,
confirming that the precursor properties do not depend on the choice
of exact threshold values.

Several crystalline symmetries appear as we further analyze the
local order of nuclei and precursors. For nuclei, three distinct
symmetries -- BCC, HCP, and FCC -- are illustrated by the
$W_4$-$q_4$ plot in Fig.1E. Clearly, nuclei at the early stage are
composed by various components \cite{gasserscience01}, which evolve
into one stable solid during much later development. Moreover, the
same analysis on precursors reveals this symmetry differentiation
even before the solid formation: there exist BCC-like, HCP-like and
FCC-like precursors whose local structures are similar to the
corresponding nuclei, as shown in Fig.1F and Fig.SI-4C. For the
first time, the experiment reveals multiple symmetries emerging in
precursors within the liquid stage, which behave as the ``seeds'' of
the multiple metastable solids and thus account for their kinetic
origins.

To verify our symmetry analysis, we plot the radial distribution
function, $g(r)$, for the three types of nuclei symmetries in
Fig.1G: all peaks match precisely with the ideal crystals,
confirming the validity of our bond order analysis (note that HCP
and FCC have similar $g(r)$ as expected). Moreover, we find a good
match in the main peaks between the precursors (Fig.1H) and the
corresponding nuclei (Fig.1G), confirming their underlying
structural connection. Because of this structural similarity, the
existence of precursors at the liquid-solid interface may
dramatically reduce the interfacial tension and promote the rate of
nucleation\cite{gasserscience01, tanakapnas}. A small surface
tension is also consistent with the ramified nuclei profile observed
in the early crystallization stage both in this experiment (see
Fig.1A and Fig.1B as an example) and in the previous study
\cite{gasserscience01}.

In addition, we clarify one important point: the local order in
precursors revealed by us is quite short-ranged, typically extending
only to the first shell (the central particle plus its neighbors).
When longer ranges are involved, however, the bulk phase of
precursor clusters is quite amorphous, due to the mixing and
interfering of multiple components (see Fig.2A). In comparison to
the previous studies, our precursor definition is similar to the one
by Kawasaki \emph{et al} \cite{tanakapnas}; but is probably only
part of the precursors defined in some other studies \cite{Tothprl,
Schillingprl, SchopePRL, Lutskoprl}, which may also contain a truly
amorphous fraction. To study this truly amorphous fraction, dynamic
investigations regarding the lifetime of dense amorphous clusters
would be needed. In conclusion, the precursors in our study exhibit
multiple local orders at short-range, but are amorphous at
intermediate and long ranges (see more details in SI, session IIC,
page 10-15).\\

\noindent\textbf{Kinetic pathways during precursor-to-solid conversion}\\
Our experiment confirms the existence of precursors, which exhibit
various symmetries and convert into different metastable solids.
However, the symmetry transformation during the precursor-to-solid
conversion requires further illumination. To visualize the process,
we color the three symmetries differently and demonstrate their
evolutions in Fig.2A-C (movie S2): at the very beginning in Fig.2A,
the HCP symmetry (nuclei + precursors, purple spheres) dominates the
other two, while the BCC symmetry (red spheres) catches up around
the critical nuclei size in Fig.2B, and later dominates the system
(see Fig.2C). This particular sample eventually develops into a BCC
crystal.

The system evolves from HCP-dominant to BCC-dominant, indicating
transformations across different symmetries. To figure out the
transformation pathways, we track the relative fractions of
different symmetries with respect to time, for both precursors and
nuclei. The typical result for a BCC-stable system is demonstrated
in Fig.2D: for all three symmetries, the precursor (open symbols)
and nuclei (close symbols) curves are roughly parallel,
demonstrating their close correlation. However, a gap also exists
between the two, revealing cross-symmetry transformations during the
precursor-to-solid conversion. More specifically, in the HCP panel
the precursor curve is higher than the nuclei one, indicating an
extra fraction of HCP-like precursors that must convert into other
types of nuclei. Correspondingly, in BCC and FCC panels excess
fractions of nuclei (instead of precursors) are found, reconfirming
the conversion from HCP-like precursors. Therefore our data indicate
two cross-symmetry pathways during the precursor-to-solid
conversion: (1) HCP-like precursors to BCC nuclei and (2) HCP-like
precursors to FCC nuclei. The same kinetic pathways are also
observed in a typical FCC-stable system, as shown in Fig.2E. In
addition, hard-sphere systems exhibit the same behavior (see
Fig.SI-12 in SI) as well, demonstrating the general validity of the
two cross-symmetry pathways.

To visualize the pathways directly, we illustrate all possible
conversions between different types of precursors and nuclei in the
$\overline{q}_6-q_4$ plane in Fig.2F: the $\overline{q}_6$ values
are used for distinguishing nuclei from precursors; while the $q_4$
values are used for distinguishing different symmetries. The lower
cluster indicates precursors, the upper cluster represents nuclei,
and the connection in between is from the particles caught during
the precursor-to-nuclei conversion which directly visualizes the
kinetic pathway. Clearly, a pathway always exists for a conversion
within the same symmetry as demonstrated by the three corresponding
images in Fig.2F (upper-left to lower-right diagonal), which
verifies the previous simulation \cite{tanakasr}. More
interestingly, cross-symmetry pathways previously never reported
also appear. Two strong pathways from HCP-precursors to BCC-nuclei
and FCC-nuclei are revealed in the second row, consistent with
Fig.2D and Fig.2E. At the same time, two weak pathways from
BCC-precursors and FCC-precursors to HCP-nuclei are also found,
although overwhelmed by the effect from the two strong ones just
mentioned. For the interaction between BCC and FCC symmetries, there
is essentially no pathway during the conversion. These kinetic
pathways are universally observed in all our soft-repulsive systems.

In soft-repulsive systems, we observe dominant HCP-like precursors
initially \cite{tanakapnas, tanakasr}, which can subsequently
convert into BCC and FCC nuclei via two cross-symmetry pathways.
However, the pathway to BCC-nuclei is much more significant, as
demonstrated by its much stronger magnitude (see Fig.2F, 2nd row,
the leftmost image compared with the rightmost image).
Correspondingly, BCC-symmetry dominates the early metastable solid
phase in both our soft-repulsive samples and the systems previously
studied
\cite{alexprl78,woldeprl95,woldejcp96,shenprl96,frenkeljpcm02,Moroniprl05}.
Therefore, we propose that the dominant cross-symmetry pathway
probably leads to the dominant metastable solid in early
crystallization. To further test this scenario, we perform the same
measurements in hard-sphere systems, whose dominant metastable solid
is RHCP (mixture of FCC and HCP) instead of BCC
\cite{gasserscience01, Auernature01}. As expected, the dominant
pathway changes accordingly: the HCP-like precursors now mainly
transfer into FCC-nuclei instead of BCC, as FCC is one important
component of the RHCP structure (see Fig.SI-12 in SI). Both the
soft-repulsive and hard-sphere experiments are consistent with our
newly proposed scenario, which explains the kinetic origin of the
\emph{dominant} metastable solid in early crystallization.

We summarize our crystallization kinetics in Fig.2G. As temperature
drops below the melting point, precursors emerge out of liquid, with
the HCP-like component dominating the BCC-like and the FCC-like
ones. These precursors then convert into metastable solids. Besides
the conversion within the same symmetry, interestingly,
cross-symmetry pathways also exist: there are two major ones from
HCP-like precursors to BCC and FCC nuclei, and two minor ones from
BCC-like and FCC-like precursors to HCP nuclei, as represented by
the solid and dashed lines respectively. The effect of major
pathways overwhelm the effect of minor ones. Eventually all
metastable solids evolve into the stable solid \cite{note1}.\\

\noindent\textbf{Density-symmetry correlation}\\
In the process of crystallization, the establishment of crystalline
order is naturally accompanied by density change. We illustrate
their correlation by directly measuring the variation of local
density with respect to the development of local order. The local
density, $\rho$, is determined by the Voronoi diagram; and the local
order is quantified by the solid bond number, $\xi$ (see SI for the
exact definition). Typically a larger $\xi$ indicates a better
crystalline order. Good statistics is obtained from measurements
over $10^6$ particles in the early crystallization stage (solid
fraction $\sim5\%$). At each solid bond number, we obtain the
average local density and illustrate their correlation in the
$\rho$-$\xi$ plot.

A typical $\rho$-$\xi$ dependence is shown by the solid symbols in
Fig.3A for a BCC-stable system. Clearly there exists a density
plateau over a broad range of local order, $3\le\xi\le10$,
indicating very little density change against a significant
development of order. In particular, across the hatched area of
liquid-solid boundary ($6\le\xi\le8$), $\rho$ does not experience
any abrupt increase, in sharp contrast to the density jump described
in the classical nucleation theory. Significant densification only
occurs at either quite early ($\xi < 3$) or rather late ($\xi > 10$)
stages of local order construction, and corresponds respectively to
the initial formation of precursors and the final perfection of
local order. We note that particles with $3\le\xi\le6$ largely
overlap with the precursor particles defined previously (see their
similar $g(r)$ in Fig.SI-6). Consequently, our experiment reveals a
rather complex three-stage correlation for density and symmetry
development: initially they grow simultaneously during the formation
of precursors ($\xi < 3$), followed by a decoupling plateau
throughout precursor-to-solid conversion ($3\le\xi\le10$), and then
reappears the simultaneous growth for the final perfection of local
order ($\xi > 10$). This three-stage trend has never been reported
previously and requires further theoretical explanation.

Different symmetries and structures emerge as early as in the liquid
stage, and therefore we illustrate their density variations
separately as well. We can divide the densities of most particles
($>98\%$) into two branches: the 14-neighbor branch (BCC-like) and
the 12-neighbor branch (FCC-like and HCP-like), as shown by the open
symbols in Fig.3A. The two branches exhibit a large difference at
low $\xi$, which diminishes as crystalline order increases.
Moreover, before the order is fully constructed, the 14-neighbor
branch is always below the 12-neighbor one, indicating a lower
density for the BCC-like structure. Since we observe a major pathway
from HCP-like precursors to BCC nuclei in the soft-repulsive
systems, it indicates a conversion from the high-density branch to
the low-density one. This conversion apparently lowers the
high-density branch, raises the low-density one, while keeps the
average density largely unchanged, as revealed by the plateau of
solid symbols. Similar results are also observed in the FCC-stable
systems shown in Fig.3B, except with two minor differences: (1)
there is a sharper density increase after the plateau, and (2) the
fully developed FCC crystal (i.e., the end point of upper branch) is
denser than the BCC counterpart (i.e., the end point of lower
branch) while the opposite is true in Fig.3A.

Over a broad range of order construction, a decoupling plateau
between density variation and symmetry development is observed,
especially at the liquid-to-solid transition boundary
($6\le\xi\le8$). Such discrepancy with the classical picture is even
more apparent when we compare the spatial distributions of $\xi$ and
$\rho$ in the early crystallization stage: very little correlation
is found between the nucleation events indicated by $\xi$ in Fig.3C
and the dense regions specified by $\rho$ in Fig.3D. The densest
regions ($\rho>1.02$, orange areas) are significantly denser than
the nuclei ($\rho\sim1.01$), and appear mostly at disordered
locations. At the same time, the local density fluctuations in both
space and time are more than $5\%$, much larger than the density
mismatch between liquid and nuclei ($1\%\sim 2\%$). To further test
this result, we plot the spatial distribution of local bond order,
$\overline{q}_6$, in Fig.3E, and again find very little correlation
with the density plot in Fig.3D. These results unambiguously
demonstrate that nucleation events \emph{do not} start from the
densest regions, in sharp contrast to the conventional picture of
nucleation. However, the similarity between Fig.3C and Fig.3E
indicates a strong correlation between the nucleation events and the
local bond order\cite{tanakapnas, tanakasr}.\\

\noindent\textbf{Discussions}\\
One important question remains unclear: why the HCP-like precursors
dominate initially? We believe it is due to the common feature of
tetrahedra clusters adjacent by faces, which widely appear in both
the disordered liquid \cite{Medvedevprl2007}and the HCP-like
precursors. This structural similarity makes the transformation to
HCP-like precursors a major first step. Subsequently, the local
HCP-like symmetry converts into other symmetries via slight
deformations: a shear parallel to the hexagonal plane can result in
FCC symmetry while an in-plane compression (or stretch) may lead to
BCC structure. Both approaches only involve locally small movements
with little energy cost, while they produce significant changes in
symmetry. The underlying inhomogeneous stresses may come from the
surface tension of ramified liquid-solid interfaces, as supported by
the observation that symmetry conversion mainly occurs at this
interface. However this picture is largely qualitative and calls for
further study.

Moreover, the density fluctuations at single particle level exceed
the liquid-solid density mismatch substantially, implying that local
structures can easily fluctuate in and out of the solid density.
Combining with the observation that nucleation rarely starts from
the densest regions, we suspect that density fluctuation may not be
the main driving factor for our crystallization process; instead the
local bond order fluctuation could be a reasonable
candidate\cite{tanakasr}. Illustrating these emerging questions
should shed new light on the conventional picture of
crystallization.

\noindent\textbf{methods}

Our system is made up of NBD dyed poly(methyl methacrylate) (PMMA)
colloids with diameter $\sigma=2.2 \mu m$ and polydispersity less
than 2.5$\%$. The colloids are suspended in a mixture of non-polar
and weakly-polar solvents, which closely matches both the refractive
index and the density of particles (see Experimental Details and
Phase Diagram in the Supplementary Information (SI) for more
details). The particles are charge-stabilized in the solvent, with
the weakly-screened electric repulsion causing Wigner crystals at
low concentrations \cite{LeunissenNature2005, Pengprl2012}. To
ensure soft-repulsive interaction, we fix the concentration of all
samples at 17\% which avoids direct contacts among particles. By
adjusting the volume ratio between the non-polar to weakly-polar
solvents \cite{LeunissenNature2005}, we can tune the Debye screening
length, $\kappa^{-1}$, and the surface charge, $Z$ \cite{Hynninen}.
This leads to different stable solid states: at long screening
lengths, the BCC symmetry is most stable; while at short screening
lengths, the most stable symmetry becomes FCC. We measure the phase
diagram with various samples (see Fig.SI-1A in SI), and show the
detailed analysis of two typical samples, BCC-stable
($\kappa^{-1}\sim$960 nm) and FCC-stable ($\kappa^{-1}\sim$520 nm),
respectively. The adjustable stable states offer the opportunity to
study various precursor-to-solid(metastable) kinetic pathways. We
shear melt colloidal crystals and record the re-crystallization
process immediately after agitation, with a Leica Sp5 confocal
microscope scanning in three dimensions with respect to time. We can
estimate the degree of supercooling below the melting temperature,
$\Delta T = T_m - T$, for all our solid samples in the phase
diagram. Two different approaches are applied: the distance from the
melting line yields $\Delta T = 0.2 T_m\sim0.44 T_{m}$; and the
Lindemann parameter approach \cite{Meijer,zahnprl00} gives $\Delta T
= 0.2T_m\sim 0.4 T_{m}$. The two independent estimates agree with
each other within the experimental accuracy.

\noindent\textbf{Correspondence and requests}: Correspondence and requests for materials should be
addressed to L. X.

\noindent\textbf{Acknowledgements}: P. T. and L. X. are supported by the Research Grants Council of Hong Kong (GRF grant CUHK404211,
ECS grant CUHK404912, CUHK Direct Grant 4053021), and N. X. is
supported by National Natural Science Foundation of China (No.
91027001 and 11074228), National Basic Research Program of China
(973 Program No. 2012CB821500), CAS 100-Talent Program (No.
2030020004), and Fundamental Research Funds for the Central
Universities (No. 2340000034). We thank Hajime Tanaka and Eli
Sloutskin for helpful discussions, and Andrew Schofield for
providing the particles.

\noindent\textbf{Author contributions}: P. T. and L. X. conceived and designed the
experiments, P. T. performed the experiments, P. T., N. X. and L. X.
analyzed the data, P. T. developed the new approach of local bond
order analysis, P. T. and L. X. wrote the paper.


\begin{thebibliography}{99}
\bibitem{note1} We find the dominant solid phase in our experimets is BCC($\lambda\sim$960 nm) or RHCP($\lambda\sim$520 nm) after several days.
\end{thebibliography}



\begin{figure}
\includegraphics[width=4.3 in]{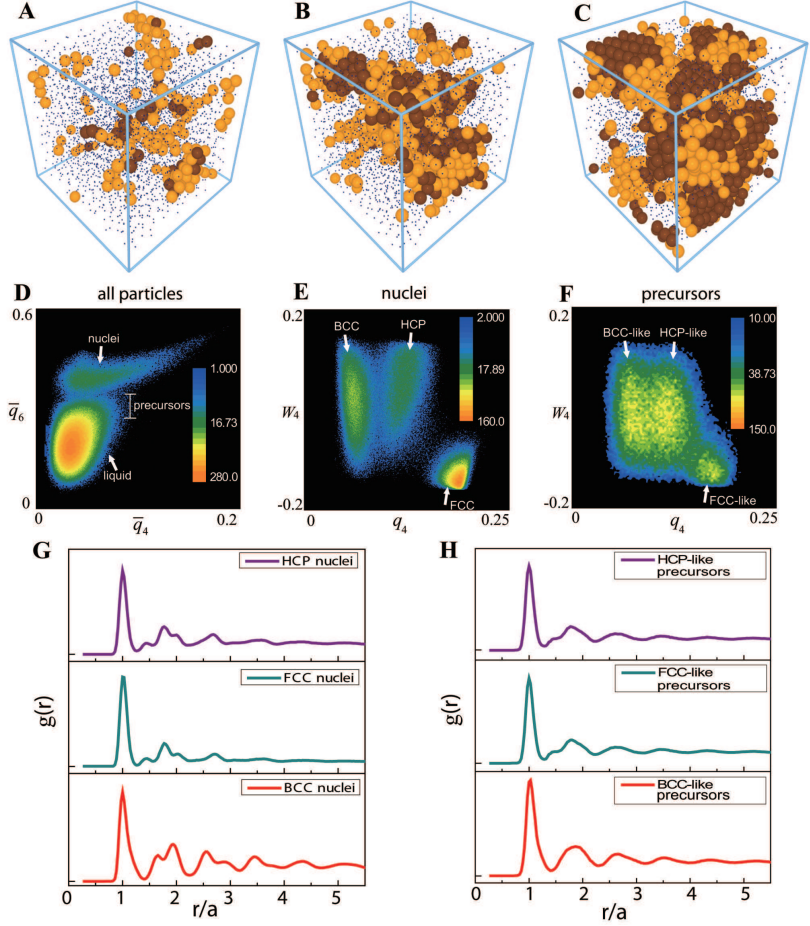}
\caption{\textbf{Structures of nuclei and precursors in early
crystallization}. \textbf{A-C}, precursor mediated crystallization
in a BCC-stable ($\kappa^{-1}=960$nm) system. The dark-brown spheres
represent nuclei particles while the light-brown ones indicate
relatively-ordered liquid particles we define as precursors. The
blue dots are normal liquid particles. At the beginning of
nucleation (\textbf{A}, $\phi_{nuclei}$ = 1.9\%, $\phi_{precursor}$
= 5.6\%, $t$ = 240s), nuclei embryos emerge from clusters of
precursors. During the subsequent development around critical
(\textbf{B}, $\phi_{nuclei}$ = 7.1\%, $\phi_{precursor}$ = 11.6\%,
$t$ = 2640s) and post-critical (\textbf{C}, $\phi_{nuclei}$ =
21.4\%, $\phi_{precursor}$ = 20.4\%, $t$=6480s) nuclei sizes,
precursors are continuously created around nuclei. \textbf{D},
number distribution of all particles in the
$\overline{q}_{6}-\overline{q}_{4}$ plane in the early
crystallization (solid fraction $\sim 5\%$). The upper group with
higher $\overline{q}_6$ is composed by nuclei particles; and the
lower group is composed by liquid particles. Within the liquid
group, relatively-ordered particles with $\overline{q}_6>0.27$ are
defined as precursors. \textbf{E}, number distribution of nuclei in
the $W_{4}$-$q_{4}$ plane reveals three meta-stable crystalline
symmetries: BCC, HCP and FCC. \textbf{F}, $W_{4}$-$q_{4}$ plot of
precursors reveals BCC-like, HCP-like and FCC-like precursors in the
liquid stage. \textbf{G} and \textbf{H}, radial distribution
function for the three types of nuclei and precursor symmetries. $r$
is re-normalized by the average particle distance, $a$. A good match
in the main peaks between \textbf{G} and \textbf{H} suggests a
structural similarity between precursors and nuclei.}
\end{figure}
\clearpage
\newpage

\begin{figure}
\includegraphics[width=5 in]{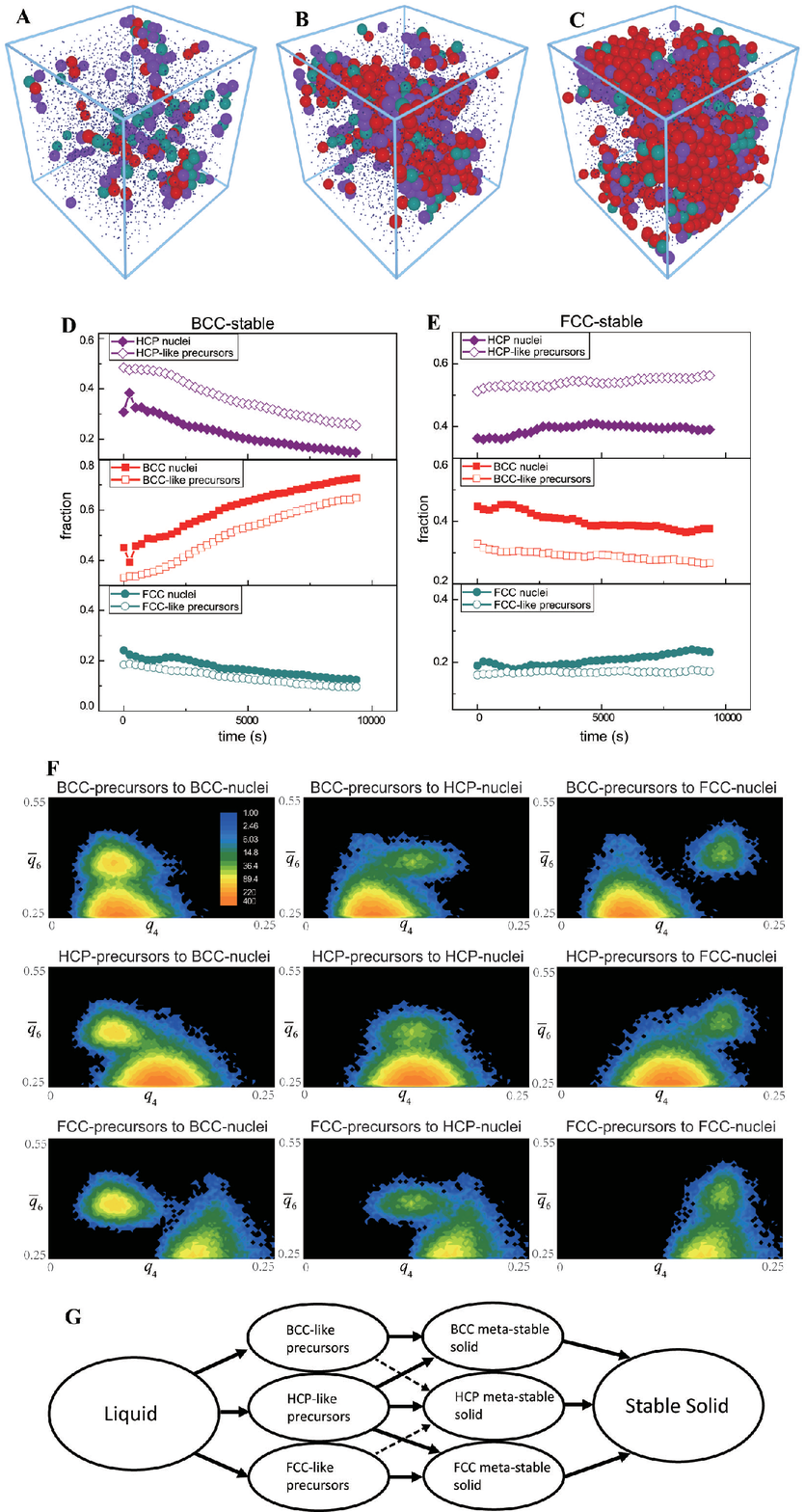}
\end{figure}
\clearpage
\newpage
\begin{figure}
\caption{\noindent Fig.~2. \textbf{Kinetic pathways during crystallization}.
\textbf{A-C}, Time evolution of different symmetries. The purple
spheres indicate HCP nuclei and HCP-like precursors, the red spheres
represent BCC nuclei and BCC-like precursors, and the green spheres
demonstrate FCC nuclei and FCC-like precursors. The three instants
are identical to the ones in Fig.1A-C. The HCP-symmetry (purple)
initially dominates, while the BCC-symmetry (red) takes over
afterwards. \textbf{D}, relative fractions of precursors (open
symbols) and nuclei (close symbols) for every symmetry in a
BCC-stable ($\kappa^{-1}=960$nm) system. The precursor and nuclei
curves are mostly parallel, indicating their strong correlation. In
the HCP panel, the precursor curve is above the nuclei one; while in
BCC and FCC panels, the precursor curve is below the nuclei one.
This indicates two kinetic pathways from the HCP-like precursors to
BCC and FCC nuclei. The critical nuclei size is reached around
$t=2640s$. \textbf{E}, the same pathways are also observed in an
FCC-stable ($\kappa^{-1}=520$nm) system. The critical nuclei size is
reached around $t=2400s$. \textbf{F}, direct visualization of
kinetic pathways in $\overline{q}_6-q_4$ plane. Each image
illustrates the conversion from one type of precursors to one type
of nuclei. The lower cluster indicates precursors, the upper cluster
represents nuclei, and the connection in between visualizes the
pathway. For conversions within the same symmetry, a pathway always
exists. For cross-symmetry conversions, there are two major pathways
from HCP-precursors to BCC and FCC nuclei and two weak pathways from
BCC and FCC precursors to HCP-nuclei. There is essentially no
pathway between the FCC and BCC symmetries. \textbf{G}, summary of
kinetics during the precursor-mediated crystallization. Solid lines
indicate the major pathways, while dashed lines represent the weak
pathways.}
\end{figure}

\clearpage
\newpage
\begin{figure}
\includegraphics[width=4.5 in]{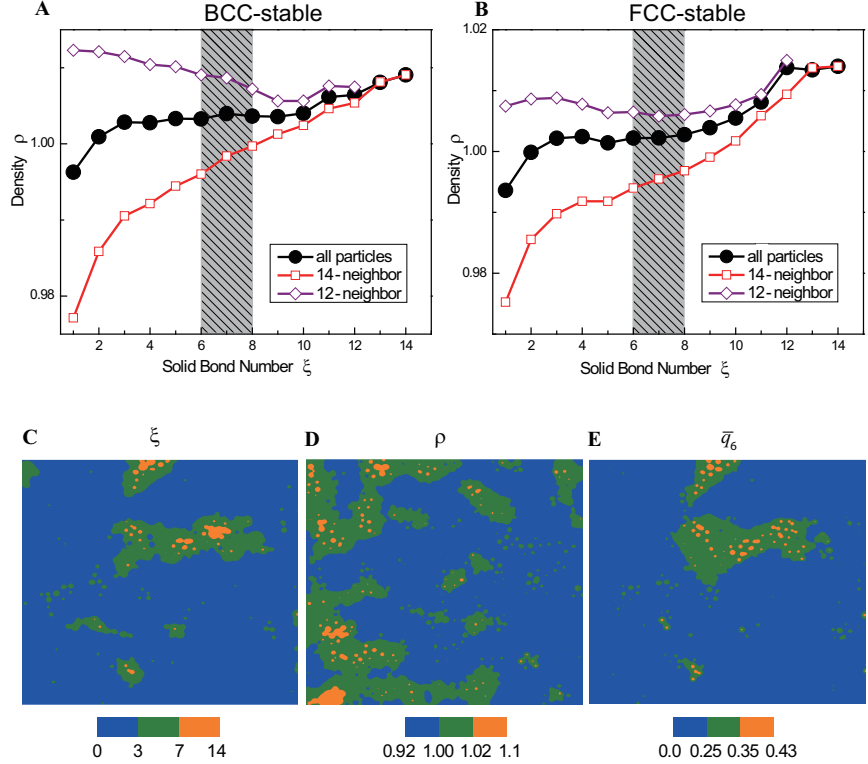}
\caption{\textbf{Local density evolution throughout
crystallization}. \textbf{A}, the local density, $\rho$, versus the
solid bond number, $\xi$, in a BCC-stable($\kappa^{-1}=960$nm)
system. The solid symbols represent the average local density as a
function of the local order development. As $\xi$ increases, $\rho$
first experiences a rapid increase which corresponds to the
precursor formation, followed by a broad plateau ($3<\xi<10$) which
corresponds to the precursor-to-solid conversion. In particular, no
density jump is observed at the hatched area of liquid-solid
transition boundary, in sharp contrast to the CNT description. The
second increase occurs at $\xi>10$ as the local order undergoes the
final perfection in nuclei. The open symbols demonstrate the
14-neighbor branch (BCC) and the 12-neighbor branch (HCP and FCC).
At low $\xi$ there exists a large gap between the two branches,
which diminishes with the increase of order. Before the order is
fully constructed, the 14-neighbor branch is always below the
12-neighbor one, indicating the lower density of BCC structure.
However, eventually the 14-bond (BCC) crystal becomes denser than
the 12-bond (FCC and HCP) crystal. \textbf{B}, the same measurements
in an FCC-stable ($\kappa^{-1}=520$nm) system. A similar three-stage
correlation is observed, but the end point of the 12-neighbor branch
is now higher than the end point of the 14-neighbor branch.
\textbf{C}-\textbf{E}, a snapshot of the spatial distributions for
the solid bond number ($\xi$), local density ($\rho$) and the local
bond order parameter ($\overline q_6$). The data are taken in a
5$\mu$m thick slice at the early crystallization stage. Clearly the
density plot has a poor correlation with the other two, indicating
that the nucleation events (regions with high $\xi$ and $\overline
q_6$ values) rarely start from the densest regions.}
\end{figure}

\end{document}